\documentclass[Royal,sageh,times]{sagej} 

\usepackage{moreverb,url}
\usepackage[colorlinks,bookmarksopen,bookmarksnumbered,citecolor=red,urlcolor=red]{hyperref}

\newcommand\BibTeX{{\rmfamily B\kern-.05em \textsc{i\kern-.025em b}\kern-.08em
T\kern-.1667em\lower.7ex\hbox{E}\kern-.125emX}}

\def\volumeyear{2022}

\begin{document}

\runninghead{Folco et al.}

\title{Data-driven micromobility network planning for demand and safety}

\author{Pietro Folco\affilnum{1}, Laetitia Gauvin\affilnum{1}, Michele Tizzoni\affilnum{1,2}, and Michael Szell\affilnum{1,3,4}}

\affiliation{
\affilnum{1}ISI Foundation, Via Chisola 5, 10126 Torino, Italy
\affilnum{2}University of Trento, 38122 Trento, Italy
\affilnum{3}IT University of Copenhagen, Rued Langgaards Vej 7, 2300 K{\o}benhavn, Denmark
\affilnum{4}Complexity Science Hub Vienna, 1080 Vienna, Austria
}

\corrauth{Michael Szell, ITU Copenhagen}
\email{misz@itu.dk}

\begin{abstract}
Developing safe infrastructure for micromobility like bicycles or e-scooters is an efficient pathway towards climate-friendly, sustainable, and livable cities. However, urban micromobility infrastructure is typically planned ad-hoc and at best informed by survey data. Here we study how data of micromobility trips and crashes can shape and automatize such network planning processes. We introduce a parameter that tunes the focus between demand-based and safety-based development, and investigate systematically this tradeoff for the city of Turin. We find that a full focus on demand or safety generates different network extensions in the short term, with an optimal tradeoff in-between. In the long term our framework improves overall network quality independent of short-term focus. Thus, we show how a data-driven process can provide urban planners with automated assistance for variable short-term scenario planning while maintaining the long-term goal of a sustainable, city-spanning micromobility network.
\end{abstract}


\maketitle
\clearpage

\section{Introduction}
In the transport sector, micromobility like cycling is one of the most promising and economic solutions to the climate crisis  \citep{creutzig2015trc,gossling2019sca,lamb2021rtd,nieuwenhuijsen2020utp}. At the same time, the most ideal environments to push for more active travel are cities because most trips are intraurban and of short length \citep{alessandretti2020shm}. However, to increase the generally low modal share of cycling and other forms of micromobility, well-designed and cohesive urban micromobility infrastructure networks are needed \citep{crow2016dmb,nieuwenhuijsen2020utp}.

Traditionally, the development of such networks follows a local, ad-hoc approach without a systematic understanding of network effects \citep{szell2021gub,natera2020dso,zhao2018bip}. Those approaches that are data-driven are typically based on survey data which provide only a rough proxy for trip demands \citep{larsen2013build,lovelace2017pct}. However, with the increased availability of high-granularity and high-frequency data sets, new data-driven approaches are emerging. On the one hand, detailed infrastructure data from crowdsourced platforms like OpenStreetMap (OSM) and their eased access via novel computational tools \citep{boeing2017osmnx} enable scientific inquiry of structure-based strategies and limitations to bicycle network growth \citep{szell2021gub,natera2020dso,vybornova2022adm}. These approaches place value on structural network properties while sacrificing specifity such as concrete demand models. The idea is that independently of the current demand patterns, eventually a cohesive, city-spanning network should be established in which long-term effects such as induced demand will push the transport system into a sustainable equilibrium \citep{nelson1997ibc,lyons2016gtp,itf2021tth,oecd2021tsn}. On the other hand, modern approaches that incorporate a variety of empirical data sets can improve the estimation of flows or potential demand to better inform priorities for short-term investments \citep{olmos2020dcf}.

Here, we follow a modern data-driven approach by extending the structural-only network growth model of \cite{szell2021gub} to account for: 1) the existing bicycle network and 2) empirical micromobility data sets of e-scooter trips and bicycle crashes, collected for the city of Turin, Italy. Our approach combines the strengths of a long-term structural growth process that aims to develop a city-spanning network \citep{szell2021gub}, with flow \citep{olmos2020dcf} and crash data \citep{larsen2013build} which are important on the short-term for considering demand and avoiding deaths and injuries from crashes. Thus, our framework provides a balance on two scales: first, on the short-term task of ``putting out fires'' with surgical investments while maintaining the long-term focus on a cohesive network, and second on deciding how much to focus on one versus another data set. As we show in this paper, our framework consolidates short-term and long-term development goals, and establishes a data-informed trade-off between demand and safety. Although our main goal is to study how a network planning process could be informed by data sets in general, we also conclude with concrete development scenarios of new protected bicycle tracks for the city of Turin which plans to implement $80\,\mathrm{km}$ \citep{ecfturin} of traffic calming measures. We apply our method to the case of Turin due to data availability, but it is applicable to any city with multiple data sets to inform tradeoffs between different development goals.

\section{Data acquisition and processing}

\begin{figure*}[t]
\centering
\includegraphics{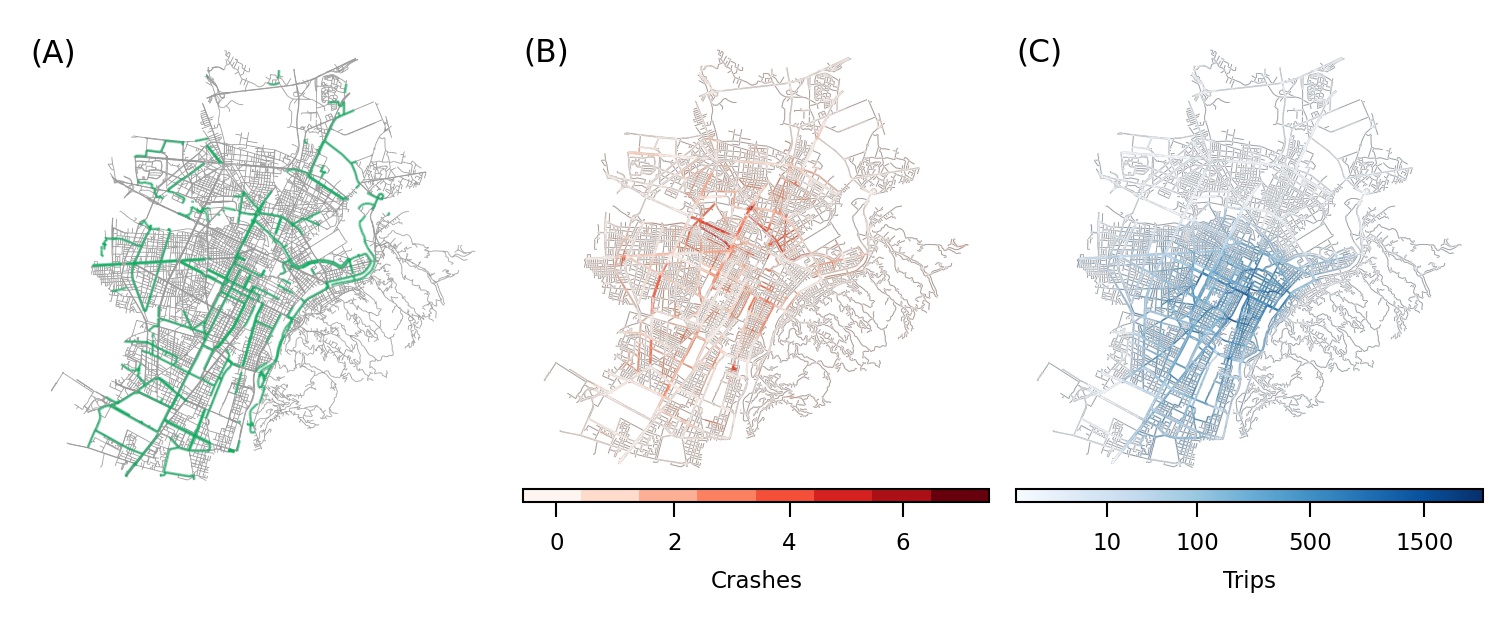}
\caption{Street and cycling infrastructure, crash and trip data for the city of Turin. \textbf{(A)} Existing cycling infrastructure is shown in green, the street network is shown in gray. \textbf{(B)} Data of road crashes that involve at least one bicycle, in the year 2019. \textbf{(C)} E-scooter trip data from May 26 2021 to October 28 2021. Trips are estimated using the locations (origins and destinations) of Bird e-scooters.} 
\label{datamaps}
\end{figure*}

\subsection{Infrastructure networks}
We downloaded existing transport infrastructure networks of Turin from OpenStreetMap (OSM) using the Python library \emph{OSMnx} \citep{boeing2017osmnx}. 
The street network and the cycling network have been downloaded and topologically simplified, where the bicycle network is the union of OSM data structures that encode on-street and off-street protected bicycle infrastructure. 
In these networks, nodes correspond to intersections, and links correspond to streets between them. 
Fig.~\ref{datamaps}A displays the street network of Turin and the existing cycling infrastructure (in green), as of July 2021. 
For additional details on the cycling network see Section S1 in the Supplementary Information.

\subsection{Bicycle crashes in Turin in 2019}
We used a dataset of bicycle crashes in Turin in 2019\footnote{Crashes in Turin (2019) available at: \url{http://aperto.comune.torino.it/dataset/elenco-incidenti-nell-anno-2019-nella-citta-di-torino}} providing a variety of information, including date and time, number and type of vehicles involved, or geographical coordinates of the crash. 
We have only considered crashes involving at least one bicycle, yielding 314 crashes in total.
Fig.~\ref{datamaps}B shows the distribution of crashes across the segments of the street network of Turin, where a crash is assigned to a street if it happened within $50\,\mathrm{m}$ of it.

\subsection{E-scooter real-time positions}
We collected real-time positions of e-scooters of the e-scooter sharing company \emph{Bird} in Turin by accessing its public API\footnote{Access instructions to the Bird API are available at: \url{https://github.com/ubahnverleih/WoBike/blob/master/Bird.md}}. 
Each query returns the position of all e-scooters that are parked within a given radius from a pair of coordinates. 
To cover the entire city, we divided it into a grid consisting of 71 grid cells of size $1506\,\mathrm{m}$ and ran a query for each centroid of the grid cells. We set the side of the cells to $1506\,\mathrm{m}$ so that all adjacent queries overlap and we do not miss data in between. Duplicates resulting from overlaps are removed. 
Time resolution of each query is $50\,\mathrm{min}$. 
We discarded cells of all the industrial, suburban, or hilly areas of the city where no e-scooters are present because parking is not allowed by the service (see Section S2 in the Supplementary Information for more details).

We collected data of the location of \emph{Bird} e-scooters from May 26, 2021 until October 28, 2021. 
The relevant features returned by each query for each e-scooter are: geographical coordinates of vehicle position, an identifying label allowing to track the movements of the vehicle, and battery charge level.
We used information on battery charge levels to identify e-scooter movements made by the company to re-locate the vehicles and charge them, and discarded them from our dataset. 
Based on the location data of e-scooters, we computed the density of e-scooters in the city and built an origin-destination (OD) matrix. 
Each origin and destination corresponds to the start and end point of an e-scooter’s movement. We define a movement if:
\begin{enumerate}
    \item[(i)] An e-scooter changes its position by at least $100\,\mathrm{m}$.
    \item[(ii)] The time between geolocation at the origin and at the destination is less than $90\,\mathrm{min}$. We add this condition because sometimes the API service has crashed and therefore some queries ($\approx 8\ \%$) are separated by more than $50\,\mathrm{min}$. We do not add movements when the queries have a separation of more than $90\,\mathrm{min}$ because the more time passes between one query and the next, the more there is the risk that a change in the position of an e-scooter is the sum of more than one movement.
\end{enumerate}

Before creating the OD matrix, we cleaned the OD data from movements made by the company: Vehicles are sometimes relocated to high-demand areas (e.g. railway stations, office areas, etc.), additionally the vehicle’s battery must be charged. To clean up the dataset from this kind of movement we used the following rule. A movement of an e-scooter is removed if one of the following two conditions is satisfied: (i) The vehicle ``disappears'' from the city and when it reappears the battery level has increased (which should happen when the company takes the vehicle to a warehouse to charge its battery), or (ii) the vehicle ``appears'' somewhere, but the battery level did not decrease with respect to the previous queries (which should happen when the company relocates the vehicle).

Finally we created a trip dataset from the OD matrix. We used each OD sample to define a \textit{trip} by computing the shortest path on the combined street and bicycle network between the origin and destination coordinates. Each calculated shortest path is a trip, yielding a dataset containing $40,\!694$ trips.
Fig.~\ref{datamaps}C shows the spatial distribution of trips over the street network of Turin. 

As of 2022, there are seven e-scooter companies operating in Turin. We have no information on how much of the \emph{Bird} data represents the overall demand and use of e-scooters for travel. However, the operational areas of all e-scooter providers are essentially overlapping.

\section{Model description}
Our network development model extends the growth framework developed by \cite{szell2021gub} by integrating existing bicycle infrastructure and data on micromobility trips and bicycle crashes. The goal of this model is a general development towards a cohesive network, but also to account for safety and demand using the crash data (proxy for safety) and the e-scooter trip data (proxy for demand). We can summarize our model in six steps:

\begin{figure*}[t]
\centering
\includegraphics{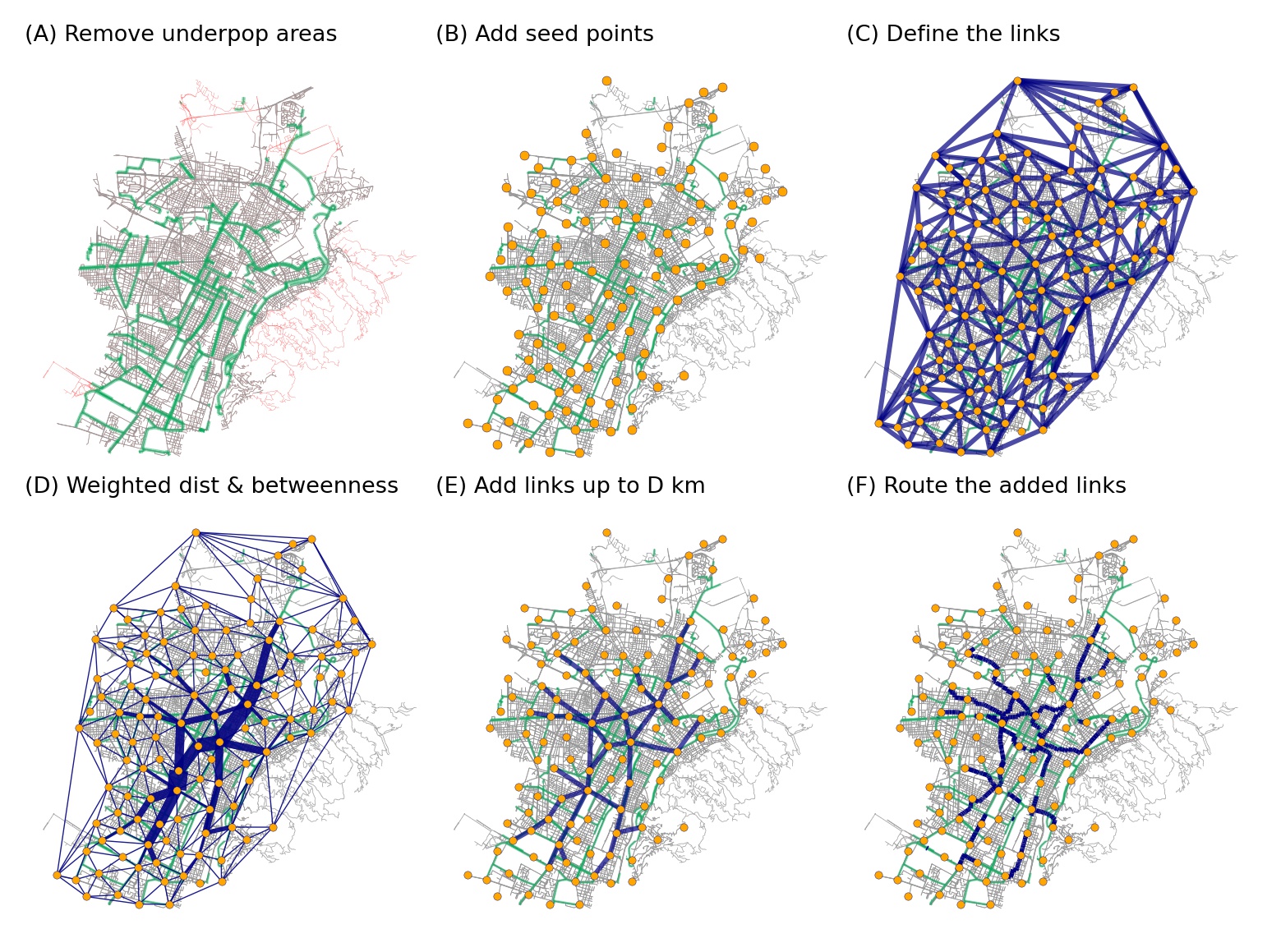}
\caption{Pipeline of the model. \textbf{(A)} Step 0: the underpopulated and hilly areas of the city (red links) are removed: no seed points will be added in these areas. \textbf{(B)} Step 1: Seed points are added following a grid structure and the existing cycle network. \textbf{(C)} Step 2: The link of the abstract network are defined. \textbf{(D)} Step 3: For each link, the weighted distance $d_{W}$ is defined, then the edge betweenness (represented with the width of the links) is calculated. \textbf{(E)} Step 4: The links are ranked by betweenness, then the top ranked links are selected up to a total of $D\,\mathrm{km}$. \textbf{(F)} Step 5: The selected links are routed on the street network. The blue links represent the solution proposed by the model, they integrate the existing cycle network (green links).} 
\label{model}
\end{figure*}

\subsubsection{Step 0 - Definition of area of study.}
The area of study we consider is the entire municipality of Turin with the exception of some underpopulated parts. 
We have chosen to remove from our analysis the underpopulated areas of the city by setting a threshold on the number of inhabitants\footnote{Number of residents in Turin (2020) by statistical areas available at: \url{http://aperto.comune.torino.it/en/dataset/popolazione-per-sesso-e-zona-statistica-2020}\\
Shapefiles of statistical areas of Turin available at: \url{http://geoportale.comune.torino.it/geodati/zip/zonestat_popolazione_residente_geo.zip}}: All the statistical areas with population density below $1,185/\mathrm{km}^2$ were removed from the analysis. 
This effectively removes the Eastern hilly part of the city which is a low demand residential area. 
We have added an exception to the above rule by including parks and green areas, even though the number of residents is obviously below the threshold as parks might include potentially safe cycling areas. 
Fig.~\ref{model}A shows the portion of the road network of Fig.~\ref{datamaps}A that was included in the study.

\subsubsection{Step 1 - Definition of seed points.}
We define a set of seed points snapped to the intersections of the entire infrastructure network (composed by the road network and the existing bicycle network). 
We also define a parameter $\delta$, as the length in meters that sets the minimum distance between two seeds.
The process of defining the seeds is the following: i) we add the seeds to the intersections of the existing infrastructure such that all seeds are at least at distance $\delta\,\mathrm{m}$ from each other, ii) we add the seeds on the intersections of the entire network from a grid of side $\delta\,\mathrm{m}$. If the distance between a seed on the existing infrastructure and a potential new seed on the grid is less than $\delta\,\mathrm{m}$, the seed is discarded (see Fig.~\ref{model}B for details). 
In the remainder of the paper, we set $\delta = 300\,\mathrm{m}$ which is reasonable with respect to the scale of the road network in Turin (length of the links, frequency of intersections). 
We also considered other values of $\delta$ in the range \mbox{200--700 $\mathrm{m}$}, and our results proved to be robust against variations in $\delta$ (see Section S3 of the Supplementary Information).

\subsubsection{Step 2 - Definition of potential links.}
Next we choose the potential links we will create between the seed points. We sort all pairs of seeds in ascending order of route distance, then we connect them stepwise with a straight link. A potential link is added if it does not cross another previously added potential link. Figure~\ref{model}C shows the result of this process. This condition, combined with the ascending orders of the pairs of seeds, follows the considerations of \cite{szell2021gub} to create a cohesive network via a greedy triangulation \citep{cardillo2006spp}, in which the closest triangle-creating pairs of seeds are connected locally. Note that this process ignores links of the existing bicycle network: if a new link would cross only edges from the existing infrastructure, it will be added nevertheless.

\cite{szell2021gub} motivate use of the greedy triangulation as optimizing investment costs \citep{cardillo2006spp}, i.e.~minimizing total length, however geometric optimization such as Delaunay triangulation could also be a possibility \citep{barthelemy2022spatial}. We compare these two methods in Section S7 in the Supplementary Information; we find high overlap between the different networks generated (Fig.~S7), qualitatively identical ranking and growth behavior of the metrics (Fig.~S8), and slightly better results for the greedy triangulation. Boundary effects are not an issue for the range of growth studied.

\subsubsection{Step 3 - Weighted distance.}
Assuming we have a budget to build $D\,\mathrm{km}$ of potential links, we have to choose which potential links to build among those defined in step 2. For this purpose, we define a weighted distance $d_W$ as the route distance adjusted with the number of crashes covered by the routed link and the number of trips passing through the routed link (to see details on how we count crashes and trips, see section S4 in the SI). We define the weighted distance $d_{W}$ for each potential link from step 2 and for each link of the existing infrastructure. For a link $e(A,B)$ we define $d_{W}(A,B)$ as follows:
\begin{align}
    d_{W}(A,B) = \alpha\ d_{\mathrm{trip}} + (1-\alpha)\ d_{\mathrm{crash}}
    \label{link_weights}
\end{align}
Where:
\begin{enumerate}
   \item[(i)] $\alpha \in [0,1]$
   
   \item[(ii)] $d_{\mathrm{trip}} = \frac{(d+1)}{(1 + 9N_{\mathrm{trip}})}$ with $ N_{\mathrm{trip}} = \frac{n_{\mathrm{trip}}}{\max(n_{\mathrm{trip}})}$
   
   \item[(iii)] $d_{\mathrm{crash}} = \frac{(d+1)}{(1 + 9N_{\mathrm{crash}})}$ with $N_{\mathrm{crash}} = \frac{n_{\mathrm{crash}}}{\max(n_{\mathrm{crash}})}$
\end{enumerate}
The variable $d$ denotes the length of $e(A,B)$ in meters, $n_{\mathrm{crash}}$ is the number of crashes per $\mathrm{km}$ covered by link $e(A,B)$ and $n_{\mathrm{trip}}$ is the number of trips per $\mathrm{km}$ passing through link $e(A,B)$. The range of the denominator of $d_{\mathrm{crash}}$ and $d_{\mathrm{trip}}$ is the interval $[1,10]$. In this way the range of variability of $d_{\mathrm{crash}}$ and $d_{\mathrm{trip}}$ is the same. We chose this definition for the denominator because it is the easiest way to set the same range of variability for both crashes and trips without adding nonlinear effects.
The weighted distance affects the edge betweenness centrality calculated for the choice of links in the next step 4: A smaller $d_{W}(A,B)$ contributes to a higher betweenness for link $e(A,B)$. In Fig.~\ref{model}D the width of potential links is proportional to their betweenness.

Using this weighting, the parameter $\alpha$ controls the influence of safety versus demand. A smaller $\alpha$ makes the model more ``safety oriented'' while a higher $\alpha$ makes the model more ``demand oriented''. The extreme values 0 and 1 define two models that use respectively only crash and trip data.

\subsubsection{Step 4 - Choose the links.}
A criterion is needed to choose a number of $\mathrm {km}$ of links equal to $D\,\mathrm{km}$ from among the potential links. Following the considerations of \cite{szell2021gub} we choose edge betweenness as a simple proxy for flow, and as an effective way to build a functional bicycle network quickly, to rank the potential links. A link with a small weighted distance is more likely to be ranked in the top positions. We add the links iteratively to the proposed solution, starting from the one with the highest betweenness, until the length of the proposed solution has reached $D\,\mathrm{km}$. The result is the network shown in Fig.~\ref{model}E.
\subsubsection{Step 5 - Routing.}
The links selected in step 4 are routed on the street network. The output of the model is a bicycle network made up of the existing infrastructure and the new $D\,\mathrm{km}$ of the proposed solution (Fig.~\ref{model}F).

\section{Evaluation metrics}
To monitor the network growth process, we generate the network up to $D=295\,\mathrm{km}$ and store a snapshot of the network every new $5\,\mathrm{km}$ of investment. 
Recently, the city of Turin announced $80\,\mathrm{km}$ of new investments in traffic calming measures \citep{ecfturin}. 
In this work we assume that the $80\,\mathrm{km}$ of announced infrastructure are equivalent to $80\,\mathrm{km}$ of new cycling paths, and we set as main target the value of $D=80\,\mathrm{km}$ of new investments.

To assess how our network growth model uses micromobility data to drive investments, we evaluate the potential improvement of two different metrics, one representing safety and the other demand: the \emph{crash coverage} and the \emph{trip coverage}.
For both metrics, we compute their potential improvement as
\begin{align}
    \pi(D) = \mu(D) - \mu(0)\,,
    \label{metric_rel}
\end{align}
where $\mu(D)$ is the metric $\mu$ calculated with $D\,\mathrm{km}$ of new added links, and $\mu(0)$ is the metric $\mu$ calculated on the existing bicycle infrastructure (i.e. when $D = 0\,\mathrm{km}$).

\subsection{Crash coverage}
To assess how many crashes (involving bicycles) would be covered by a proposed bicycle network we define a crash coverage metric. We first assume that a road segment is safer if there is a protected cycle track, so we expect a reduction in the number of crashes near it. 
Since we can not compute how many crashes would have been avoided with a cycle track, we count how many crashes are covered by the proposed cycling infrastructure, assuming that higher coverage increases road safety \citep{teschke2012route}. 
We label a crash as ``covered'' if located within $50\,\mathrm{m}$ of a bicycle infrastructure element (see Fig.~S3 in the SI for technical details). 
In this way, we can see how the number of crashes covered in the city increases as the network grows and compare these results between different settings of our model. 
Mathematically, we define the crash coverage $C$ as
\begin{align}
    C = \frac{N_{\mathrm{cov}}}{N_{\mathrm{tot}}},
    \label{crashes_coverage}
\end{align}
where $N_{\mathrm{cov}}$ is the number of crashes geolocated within $50\,\mathrm{m}$ of a bicycle infrastructure link, and $N_{\mathrm{tot}} = 314$, the total number of crashes involving bicycles in 2019.

\subsection{Trip coverage}
Parallel to the crash coverage metric we define a trip coverage metric to assess how the proposed solution fits the demand. Given $M\,\mathrm{trips}$ chosen randomly, we calculate the percentage of $\mathrm{km}$ traveled on the cycle infrastructure with respect to the total $\mathrm{km}$ traveled. Mathematically, we define the trip coverage $T$ as
\begin{align}
    T = \frac{\sum_{m} l_{\mathrm{bike}}(m)}{\sum_{m} l_{\mathrm{tot}}(m)},
    \label{trips_coverage}
\end{align}
where $l_{\mathrm{bike}}(m)$ is the number of $\mathrm{km}$ on the bicycle path of trip $m$, and $l_{\mathrm{tot}}(m)$ is the total length (in $\mathrm{km}$) of trip $m$.

In defining our metric we calculate the shortest path between origin and destination for each trip. As a baseline, we do not take into account that a user may accept a certain percentage of detour if it allows to cycle on safer infrastructure, e.g. a protected cycle track. Our way of calculating the trip (shortest path) therefore gives us the trip coverage if users accept a 0\% detour, which can be considered as a lower bound. For a more realistic analysis we also evaluated trip coverage for an accepted detour level of 25\%, which yields similar results in terms of potential improvement (see Section S5 in the Supplementary Information for details).

\begin{figure*}[t]
\centering
\includegraphics{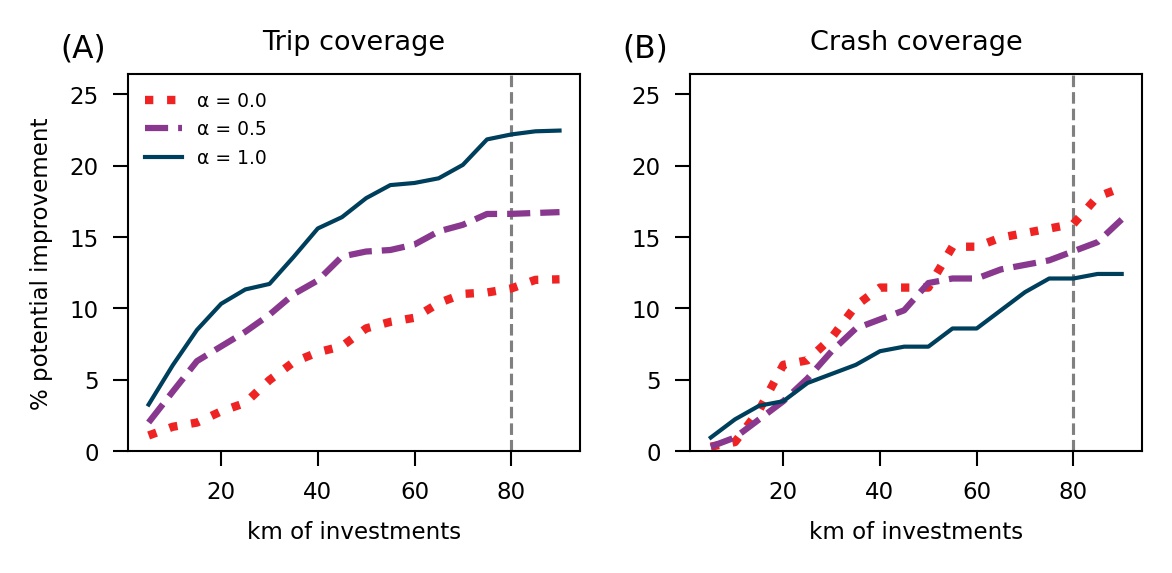}
\caption{Network metrics improvement as a function of added cycling infrastructure in kilometers. \textbf{(A)} Potential improvement of trip coverage for $\alpha = 0.0,  0.5, 1.0$. \textbf{(B)} Potential improvements of crash coverage for $\alpha = 0.0, 0.5, 1.0$. As expected, lower $\alpha$ values lead to higher improvements for a safety-focused growth; while higher $\alpha$ values correspond to higher improvements for a demand-focused growth. The dashed vertical line indicates the value $D=80~\mathrm{km}$.} 
\label{metrics_results}
\end{figure*}

\section{Results}
Using our development model, we generated networks of new cycling paths in Turin by varying the parameter $\alpha$ in the range $[0, 1]$, and for increasing values of realistic investments $D$, representing the total length in $\mathrm{km}$ to be added to the existing infrastructure. Fig.~\ref{metrics_results} shows the performance of the model in terms of potential improvement in trip coverage (Fig.~\ref{metrics_results}A) and crash coverage (Fig.~\ref{metrics_results}B), as $D \geq 5$ kilometers of new cycling paths are added to the network, for three characteristic values of $\alpha$. As expected, we obtain the largest improvement in crash coverage for $\alpha = 0$ and the largest improvement in trip coverage for $\alpha = 1$. For any value of $D$, trip coverage for $\alpha = 1$ is consistently higher than for $\alpha = 0.5$ which in turn is higher than for $\alpha = 0$. On the other hand, we see the best performance in crash coverage for $\alpha = 0$ for each $D$ (with the exception of $D \leq 10\, \mathrm{km}$ where statistical noise dominates over potential negligible improvement around $1\%$, insensitive to the choice of $\alpha$). The differences in performance as $\alpha$ varies are evident at $D = 80\, \mathrm{km}$ but also for lower values of $D$, showing that the model is reliable also at intermediate steps of the development plan. Therefore, our results confirm that the response of the model with respect to the parameter $\alpha$ provides urban planners with a choice between directing an investment plan towards the optimization of travel demand or safety simply by varying $\alpha$, in any reasonable range of investment $D$. In general, for the data at hand the potential improvement achieved by the model in trip coverage (Fig.~\ref{metrics_results}A) is higher than in crash coverage (Fig.~\ref{metrics_results}B). For instance, considering the results at $D = 80\,\mathrm{km}$, trip coverage displays a potential improvement of 22\% (corresponding to a trip coverage of $\mathrm{T} = 29\%$) for $\alpha = 1$, while crash coverage improves by only $16\ \%$ (corresponding to a crash coverage of $\mathrm{C} = 54\ \%$) for $\alpha = 0$. Such a performance gap decreases with larger values of new investments $D$, as shown in Fig.~S6 of the Supplementary Information. In the extreme case of $D=295\,\mathrm{km}$, the potential improvement of trip and crash coverage achieved by the model is about 30\%, in both cases. Tables S1, S2 and S3 of the Supplementary Information report numerical values of improvements for all metrics, and for $\alpha = 0,0.5,1$.

\begin{figure*}[t]
\centering
\includegraphics{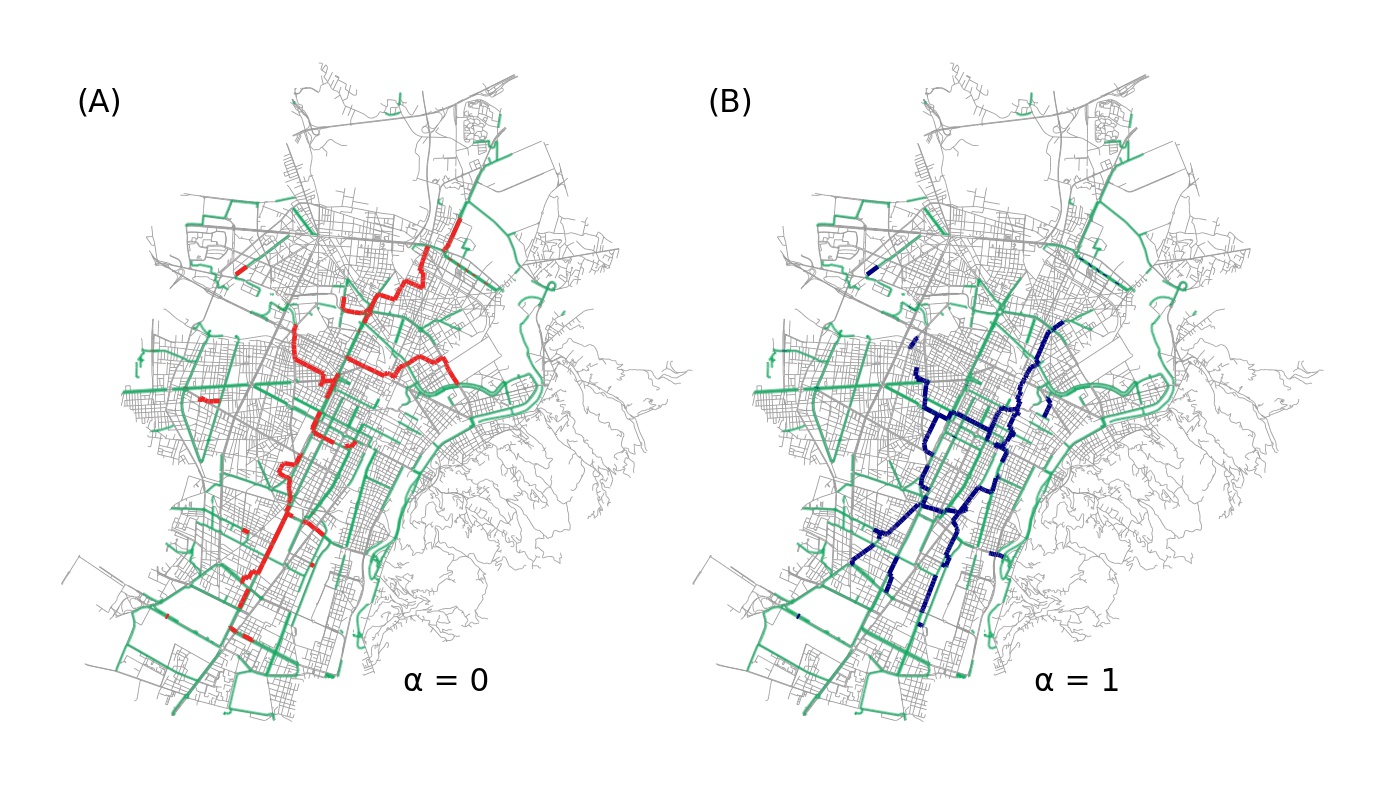}
\caption{The proposed solutions at $25\,\mathrm{km}$ of investments using the model described in section Model description. \textbf{(A)} The proposed solution for $\alpha = 0$, i.e. using only crash data. \textbf{(B)} The proposed solution for $\alpha = 1$, i.e. using only trip data.} 
\label{maps_25km}
\end{figure*}

Different choices of $\alpha$ values lead to different structures of the generated cycling network. As an example, Fig.~\ref{maps_25km} shows the first $25\,\mathrm{km}$ of new cycle infrastructure generated by the model in Turin for $\alpha = 0$ (panel A) and $\alpha = 1$ (panel B). For $\alpha = 0$, when only crash data are taken into account, new cycling paths along the North-South direction of the city are initially prioritized, while for $\alpha = 1$ most of the new links are added in the central area of the city. In particular, we can observe that for $\alpha = 1$ the model generates a cycling ring around the main train station of the city. Such different prioritizations of new links are mainly related to the spatial distribution of the crash and trip input datasets shown in Fig.~\ref{datamaps}, confirming that the model is well adapted to take into account different inputs. Crashes are more uniformly distributed throughout the city with some hotspots, while trip demand is highly concentrated in the city center, the area where most of the e-scooter trips are concentrated and the service is available (see Fig.~S1 in the Supplementary Information for a map of the area where e-scooters can be parked). Figure~\ref{maps_25km} also shows how the model integrates with the existing infrastructure by connecting its disconnected components. As expected, the number of disconnected components of the network decreases with increasing $D$, however, the trend of the number of components does not depend on $\alpha$: as $\alpha$ varies, for a given $D$, the number of components remains approximately the same (see Section S7 in the Supplementary Information for details).

So far we have presented the extreme cases of our network growth model: namely, $\alpha = 0$ and $\alpha = 1$. However, an urban planner may want to look for solutions that improve both safety and demand at the same time, striking a balance between the two. By setting intermediate values of $\alpha$, our model takes into account both crash and trip data providing a solution that balances both inputs, following the weighting approach of  Eq.~\ref{link_weights}. Thus, for a given investment in new cycling paths, $D$, it is possible to define a trade-off value of $\alpha$, $\alpha^{*}(D)$, for which both metrics, trip coverage and crash coverage, achieve the same potential improvement. By setting $\alpha = \alpha^{*}$, the model provides the most balanced solution between demand and safety. It is important to notice that an equal balance between demand and safety would correspond to $\alpha = 0.5$, in terms of weighted distance $d_{W}$, however, this does not immediately translates into an equal improvement on both metrics, due to the different spatial distributions of trips and crashes. 

\begin{figure*}[t]
\centering
\includegraphics{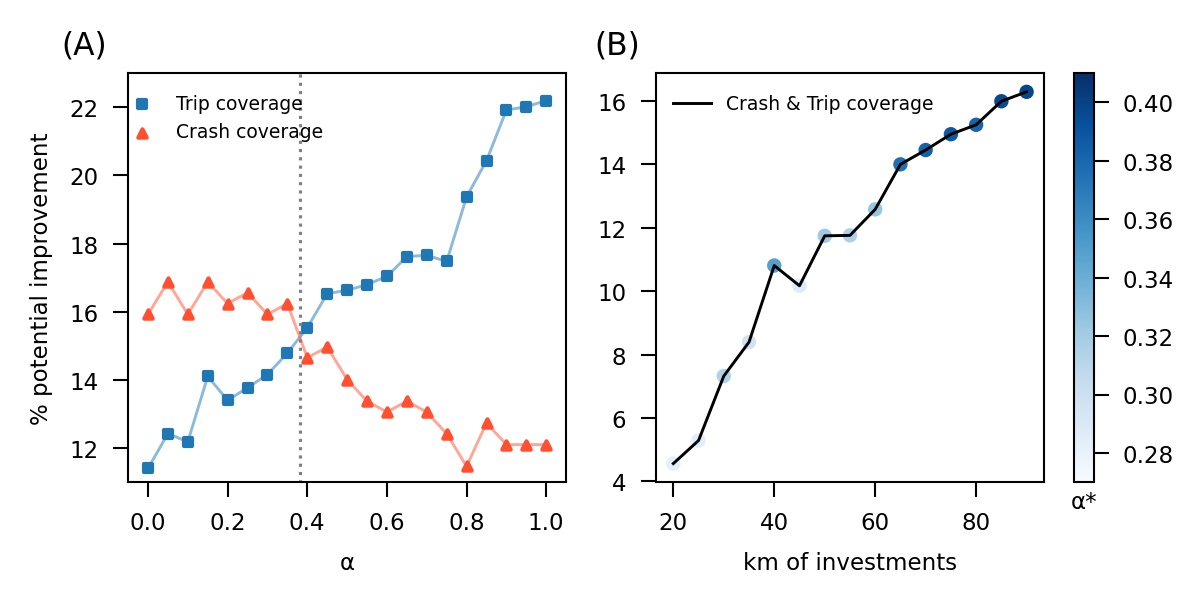}
\caption{Trade-off between crash coverage and trip coverage. \textbf{(A)} The potential improvement at $80\,\mathrm{km}$ of new investments as a function of the parameter $\alpha$. Crash coverage is decreasing with $\alpha$, while trip coverage is increasing with $\alpha$. At the value $\alpha^*$, marked by a dashed line, both metrics achieve the same relative improvement.
\textbf{(B)} The potential improvement as a function of $\mathrm{km}$ of investments, at the intersection $\alpha^*$. The colorbar represents the values of $\alpha^*$.} 
\label{tradeoff}
\end{figure*}

As an example, Fig.~\ref{tradeoff}A shows how trip coverage and crash coverage respond to the variation of $\alpha$, at $D = 80\,\mathrm{km}$. In this case, we find $\alpha^{*} = 0.38$, indicated with a dashed line on the plot. However, it must be taken into account that $\alpha^{*}$ is a quantity that depends (weakly) on $D$. Fig.~\ref{tradeoff}B shows the potential improvement in trip coverage and crash coverage at the trade-off value, for increasing $D$. By measuring $\alpha^{*}$ for varying $D$ in the range \mbox{20--90 $\mathrm{km}$}, we observed that $\alpha^{*}$ varies in a relatively small interval: $\alpha^{*} \in [0.28,0.40]$. For these values of $D$ and $\alpha^{*}$, the potential improvement in trip coverage and crash coverage increases from about $4\%$ up to $16\%$. This is a theoretical trend, since, as $D$ varies, the potential improvement is calculated for $\alpha^{*}(D)$. This dependence on investment length helps us to understand which trend is obtained at the trade-off value, but in a practical context where the final $D$ might be unknown, $\alpha$ must be set at the beginning of the process. In such a case, in Turin, a value of $\alpha$ between 0.28 and 0.40 would provide a balanced solution to generate new cycling paths that take into account both demand and safety.

\section{Discussion}
In this study, we have presented a new data-driven cycling network growth model, which aims to build up a cohesive network while incorporating multiple empirical data sets, optimizing different development targets at the same time. In particular, we developed the model to generate new cycling tracks, to be integrated into existing infrastructure, and to prioritize travel demand and cycling safety. Through one tunable parameter $\alpha$, the model generates networks by either focusing on maximizing trip coverage to satisfy existing travel demand, or by focusing on crash coverage to increase safety, or by balancing a mix of the two targets. Our work extends previous network development approaches \citep{natera2020dso, olmos2020dcf, szell2021gub} by including multiple dimensions into the model, in the effort of providing urban planners with a tool that can address different mobility needs of a city. To assist this effort, we released the code of our model publicly. 

In this paper we used the micromobility modes of e-scooters and bicycles interchangeably, being fully aware that these are not the exact same modes of transport.  They could have potentially different user profiles, usage patterns, or risks due to the novelty of e-scooters \citep{sanders2022results}. We used on one hand bicycle crashes as a proxy for micromobility safety, on the other hand e-scooter trips as a proxy for micromobility demand. It would have been preferable to have also data of bicycle trips or e-scooter crashes available -- but to the best of our knowledge those data are not collected or being made available publicly in Turin. We justify this choice as follows: First, our study does not aim to generate a final, concrete network that can be implemented 1-to-1 by the city, but we aim to study the general trade-offs between different data sets when used concurrently for data-informed network growth. From this broad perspective, it is most important that the two data sets, despite being correlated (Section~S9 and Fig.~S9 in the Supplementary Information), are different enough to produce different results, which is the case (see e.g.~Figs.~\ref{metrics_results},\ref{maps_25km},\ref{tradeoff}). Second, e-scooter and cycling crashes \citep{shah2021comparison} and their usage patterns in Turin \citep{chicco2022understanding} and elsewhere \citep{reck2021explaining} have been shown to display many similarities. Third, Italian law \citep{italianlaw} considers e-scooters and bicycles as equivalent concerning traffic rules -- they share the same network infrastructure. Fourth, in other Italian cities where both e-scooter and bicycle trip data are available, there is a strong correlation (see Section~S8 in the Supplementary Information). Lastly, the approaches to non-cycling micromobility network planning are structurally identical or very similar to bicycle network planning which we used here \citep{ignaccolo2022developing,fazio2021planning,comi2022innovative}. Many modes of micromobility are relatively understudied due to their novelty; therefore, research on the topic is highly fluid and expected to provide new insights in the near future \citep{cook2022more}.

We applied our modeling framework to the specific case of the city of Turin, but the model can be easily extended to other cities for which micromobility travel demand such as bicycle or e-scooter trips, and crash data are available. From a theoretical perspective, our results do not depend on the city under study: the model is defined to balance the trade-off between two different mobility targets, here travel demand and safety, and it allows to identify a solution that prioritizes either one through a single parameter. Urban planners could adapt the model to include different metrics to be prioritized by the proposed solution, such as the number of speed-limited streets across the cycling path, street lightning, perception of urban environment, and other measures of comfort \citep{quercia2014shortest, quercia2015smelly}. Of course, such model extensions will require the availability of geo-located data representing the metrics of interest that could be mapped onto the existing transport infrastructure. 

Our work comes with limitations and possible extensions. In particular, specific instances of the cycling networks generated by our model may be sensitive to the input data representing the target metrics. In our study, we examined the case of Turin for which we had crash data from 2019 and trip data from 2021. The crash dataset contains 314 geo-located events in total, which is large enough to allow our data-driven planning approach, but also leaves open the desire for extension to reduce possible statistical noise. Crash data are notoriously underreported, especially for vulnerable road users \citep{etsc2020,olszewski2018iso}, so a change in reporting procedures could improve the extent and statistical expressiveness of crash data ground truth. Despite general issues of reporting bias, we are not aware of concrete significant biases in the datasets we used here. Data from 2020 were also available but we did not include them into the model because urban mobility was strongly affected by COVID-19 restrictions in 2020 \citep{gauvin2021socio}, and 2020 crash data cannot be considered a representative baseline. Unfortunately, crash data before 2019 were not geo-located. A more systematic mapping of bicycle safety in Turin, through a multi-year data collection, could improve our results by providing a more precise picture of crash hotspots across the city. Also, we considered crash locations as the most immediate proxy for street safety, however other street-level features could be used to model risk, such as street width or number of intersections \citep{daraei2021data}. 

As a proxy for travel demand, we considered e-scooter trip data. The dataset under study is quite large, including more than 40 thousand trips; however, we assumed that e-scooter riders have similar mobility needs of cyclists, but this may not be always the case. Recent studies have shown that micromobility modes display different characteristics: bike sharing services are predominantly used for commuting, while e-scooter sharing is more related to recreational activities
\citep{reck2021explaining, mckenzie2019spatiotemporal}.
We do not expect such differences to significantly affect our model, since they are mainly reflected in temporal patterns of ridership rather than their spatial distribution, but a more specific disaggregation of trip demand by micromobility mode could improve our results and would be a natural extension of the model. 
Also, the e-scooter sharing service is only available within a part of the city area (see Fig. S1), therefore, trip data are limited to a portion of the city. Combining data from different e-scooter/bike sharing providers would be a solution to overcome this limitation.     

It is important to note that the relative improvement of target metrics is computed with respect to a baseline value defined at $D=0$, i.e. with the existing cycling infrastructure. Therefore, our model assumed that the existing infrastructure already satisfies trip demand and cycling safety at the baseline values of coverage. Such an assumption could be relaxed to improve model evaluation, for instance by defining different levels of trip and crash coverage, also at the baseline, depending on the type of infrastructure that already exists and that is planned. Further, our proposed approach will always result in initially centralized networks, see also \cite{szell2021gub}. However, a centralized initial solution generally reflects the typically centralized mobility flows in cities \citep{schlapfer2021universal}, and the algorithm already places the seed points in a way to cover the whole city so that eventually all points will be reached. We consider implementing a non-centralized initial solution conceptually too different and outside of the scope of this paper, but we refer to \cite{ospina2022maximal} who develop a maximum coverage design from the beginning.

Moreover, we assumed that the addition of new bicycle tracks will automatically increase cycling safety, but this is a debated issue in the literature. Although there exist empirical guidelines \citep{crow2016dmb}, there is a lack of scientific consensus on which specific type of infrastructure provides the best safety, and under which conditions it should be installed. Generally, exposure data can provide evidence for greater protection through physically separated infrastructure \citep{teschke2012route}, but safety is such a complex, yet unresolved topic requiring a deep discussion of a multitude of variables \citep{klanjcic2021iuf} or elaborate tools and analysis \citep{kondo2018bike}, that we consider it outside the scope of this work. It is however important to note that the total number of crashes is just a reflection of infrastructure safety versus route choice since we were not able to consider exposure data. For example, it might happen that unsafe roads are not classified as such, because no cyclists are using them and thus, no crashes occur there. Adequately corresponding data of bicycle trips, which we do not have unfortunately, would allow to calculate crash risk per travelled length. For this it would be important to have data over a longer observation period with a larger number of crashes, possibly assisted by state-of-the-art methods like Bayesian analysis \citep{kondo2018bike}, towards adequate statistical robustness. Nevertheless, the absolute number of crashes is important to consider for short-term investments, i.e. in the realistic scenario where the city wants to ``put out fires''.

To summarize, our approach is as an extension of the network growth model of \cite{szell2021gub}, which overcomes two previous limitations: Our new model 1) does not ignore existing cycling infrastructure but snaps seed points explicitly to existing bicycle tracks, and 2) it incorporates empirical traffic data to allow meeting short-term goals. These are two important steps away from a purely theoretical growth model towards a more realistic approach that could be useful for both underdeveloped and developed cities. By combining the goals of satisfying demand/safety and structural cohesion, we unite a short-term engineering approach of accounting for acute mobility circumstances with a long-term systemic approach of designing an accessible, sustainable mobility system \citep{oecd2021tsn}. Such a flexible, data-driven process can thus provide urban planners with a well-rounded, automated assistance for variable short-term scenario planning while maintaining the long-term goal of a sustainable, city-spanning micromobility network.

\section{Data and code availability}
Data and code to fully reproduce the results of the study are available at the repository: \url{https://github.com/pietrofolco/Data-driven_bicycle_network_planning_for_demand_and_safety}.

\begin{acks}
PF, LG, MT gratefully acknowledge the Lagrange Program of the ISI Foundation funded by CRT Foundation. MS gratefully acknowledges support by the Danish Ministry of Transport. We gratefully acknowledge the open data that this article is based on, from \url{https://www.openstreetmap.org}, copyright OpenStreetMap contributors.
\end{acks}

\bibliographystyle{SageH}
\bibliography{main}
\end{document}


\maketitle
\beginsupplement

\section{Download infrastructure networks}
We downloaded existing networks of Turin on July 12, 2021 from OpenStreetMap using \emph{OSMnx} \cite{boeing2017osmnx}. To download the street network you need to set the following parameters:
\begin{itemize}
    \item ``network\_type'': \textbf{``drive''}, ``custom\_filter'': \textbf{None}
\end{itemize}
The bicycle network is the union of several special cases. We list them here:
\begin{itemize}
    \item ``network\_type'': \textbf{``bike''}, ``custom\_filter'': [\textbf{``cycleway:left''$\sim$``track''}]
    \item ``network\_type'': \textbf{``bike''}, ``custom\_filter'': [\textbf{``cycleway:right''$\sim$``track''}]
    \item ``network\_type'': \textbf{``bike''}, ``custom\_filter'': [\textbf{``cycleway''$\sim$``track''}]
    \item ``network\_type'': \textbf{``bike''}, ``custom\_filter'': [\textbf{``highway''$\sim$``cycleway''}]
    \item ``network\_type'': \textbf{``bike''}, ``custom\_filter'': [\textbf{``bicycle\_road''}]
    \item ``network\_type'': \textbf{``all''}, ``custom\_filter'': [\textbf{``highway''$\sim$``path''}][\textbf{``bicycle''$\sim$``designated''}]
    \item ``network\_type'': \textbf{``bike''}, ``custom\_filter'': [\textbf{``cyclestreet''}]
\end{itemize}

\section{Where e-scooters can be parked}
\begin{figure*}[h]
\centering
\includegraphics{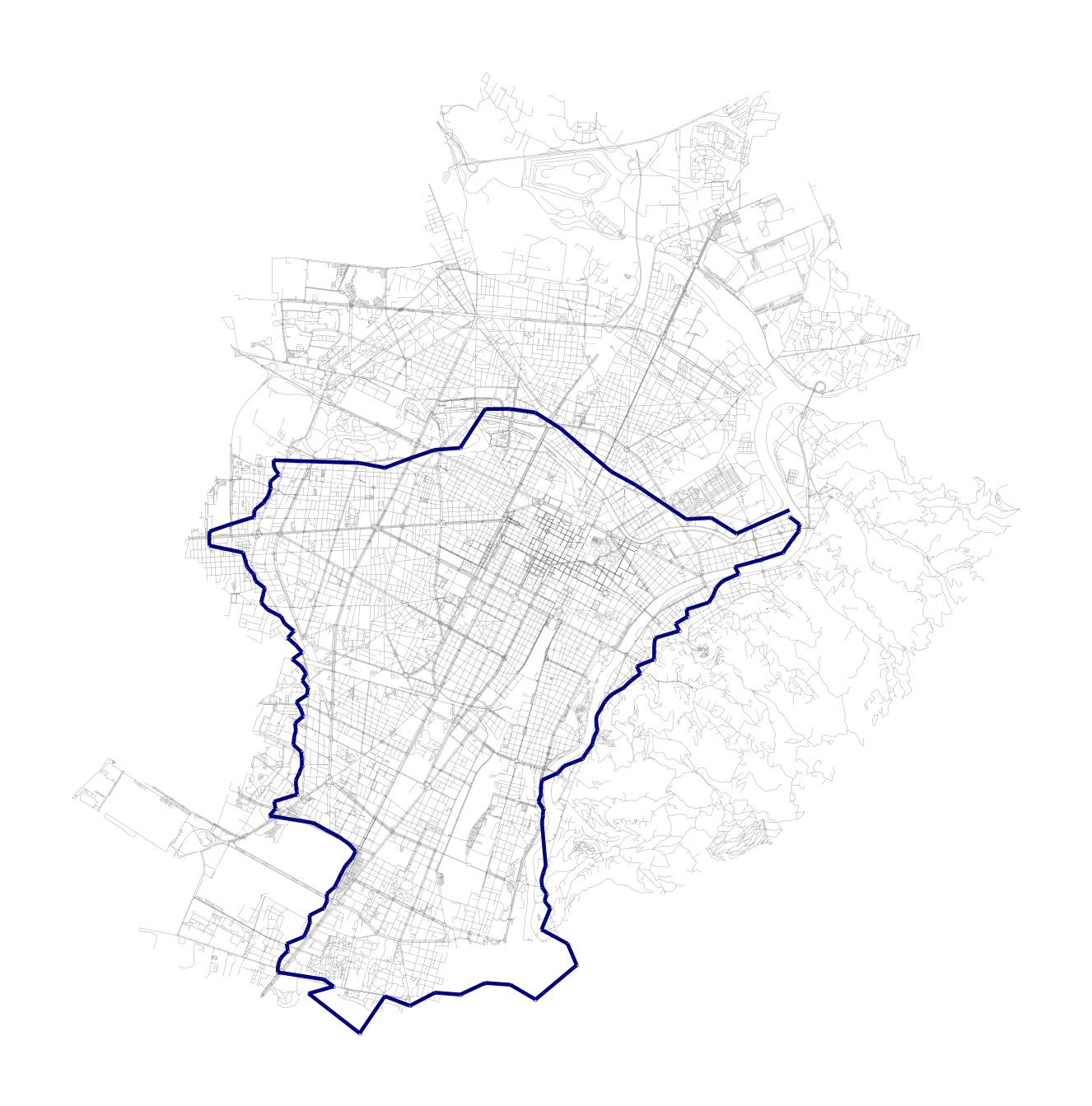}
\caption{Bird's e-scooters can be parked only inside the blue borders in Turin.} 
\label{parking_borders}
\end{figure*}

E-scooters data are available in the city area within the blue borders of Fig.~\ref{parking_borders}. In the municipality of Turin, during the entire data collection we found only e-scooters within this area. It corresponds to the area where e-scooters can be parked, as we have verified from the Bird app.

\section{Set the $\delta$ parameter}
For our main analysis we set $\delta = 300\ \mathrm{m}$, but we explored also other values: $\delta = 200\ \mathrm{m},$ $300\ \mathrm{m},$ $400\ \mathrm{m},$ $500\ \mathrm{m},$ $600\ \mathrm{m},$ $700\ \mathrm{m}$.
We reported in Fig.~\ref{SI_delta-comparison} the trends of the potential improvements for trip coverage and crash coverage ($\alpha = 0.0,\ 1.0$) in the range $D = [5,90]\ \mathrm{km};\  \mathrm{step} = 5\ \mathrm{km}$. For trip coverage we should consider the results for $\alpha = 1.0$ to understand the response of the model to $\delta$ because that is the case when only trips data are taken into account in the network growth. For the same reason we should mainly consider the results for $\alpha = 0.0$ to understand the response of the model to $\delta$ on crash coverage.
In general, we can say that the response of the model to the variation of $\delta$ is robust around the value we have chosen. For example, in Fig.~\ref{SI_delta-comparison}B the trend for $\delta = 400\ \mathrm{m}$ is very similar to the trend observed for $\delta = 300\ \mathrm{m}$, but shifted by 1-2 \%. $\delta = 200\ \mathrm{m}$ also has a similar trend to $\delta = 300\ \mathrm{m}$ especially in the first few kilometers, then we observe a less steep curve for $D > 25\  \mathrm{km}$. We observe similar trends also in Fig.~\ref{SI_delta-comparison}C, but here also  $\delta = 600\ \mathrm{m}$ revealed a curve close to that of $\delta = 300\ \mathrm{m}$.
\subsection{Too high values of $\delta$ neutralize the effect of $\alpha$.}
\begin{figure*}[h]
\centering
\includegraphics{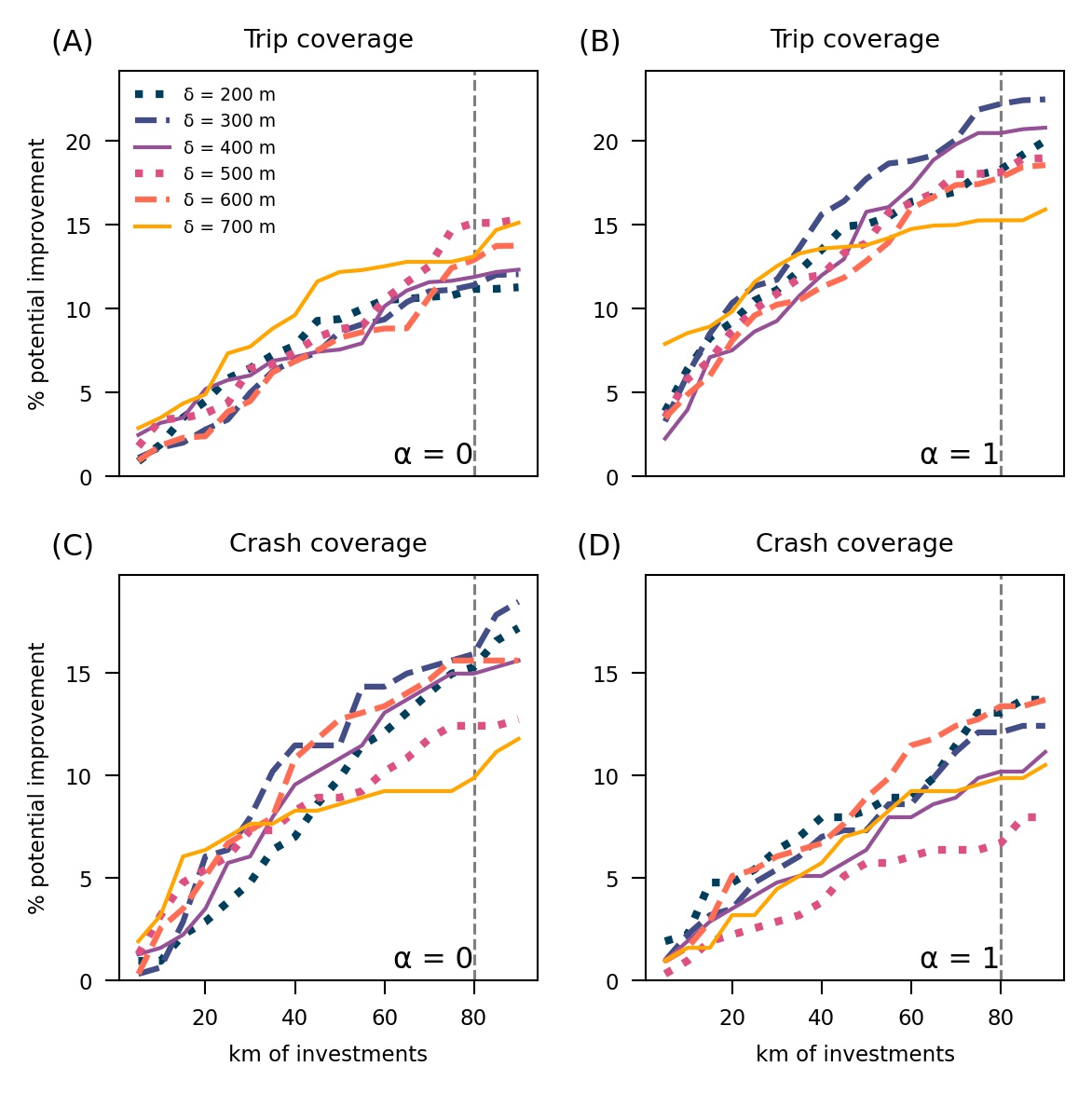}
\caption{Metrics comparison for different $\delta$ values. \textbf{(A)} Trip coverage for $\alpha = 0.0$. \textbf{(B)} Trip coverage for $\alpha = 1.0$. \textbf{(C)} Crash coverage for $\alpha = 0.0$. \textbf{(D)} Crash coverage for $\alpha = 1.0$. } 
\label{SI_delta-comparison}
\end{figure*}

From Fig.~\ref{SI_delta-comparison} we observe that for high values of $\delta$ (e.g. $\delta = 700\ \mathrm{m}$) there are no significant changes in the curve by varying $\alpha$. This is caused by the low resolution we get by setting  $\delta = 700\ \mathrm{m}$. The resolution it too low to capture inputs from our data sets. As a result, network growth is no longer driven by trip/crash data. For this reason we should use lower values of $\delta$. On the other hand, too low values of $\delta$ would increase the computation time and the performance would not improve if operating on a scale smaller than the network scale (length of the links, frequency of intersections).\\
For all these reasons we have chosen to set $\delta = 300\ \mathrm{m}$. This would also be our starting point if we extended this model to another city, but we do not exclude having to redo a fine-tuning of this parameter.

\section{Weighted distance: how crashes and trips are counted}\label{weighted_dist}
\begin{figure*}[h]
\centering
\includegraphics{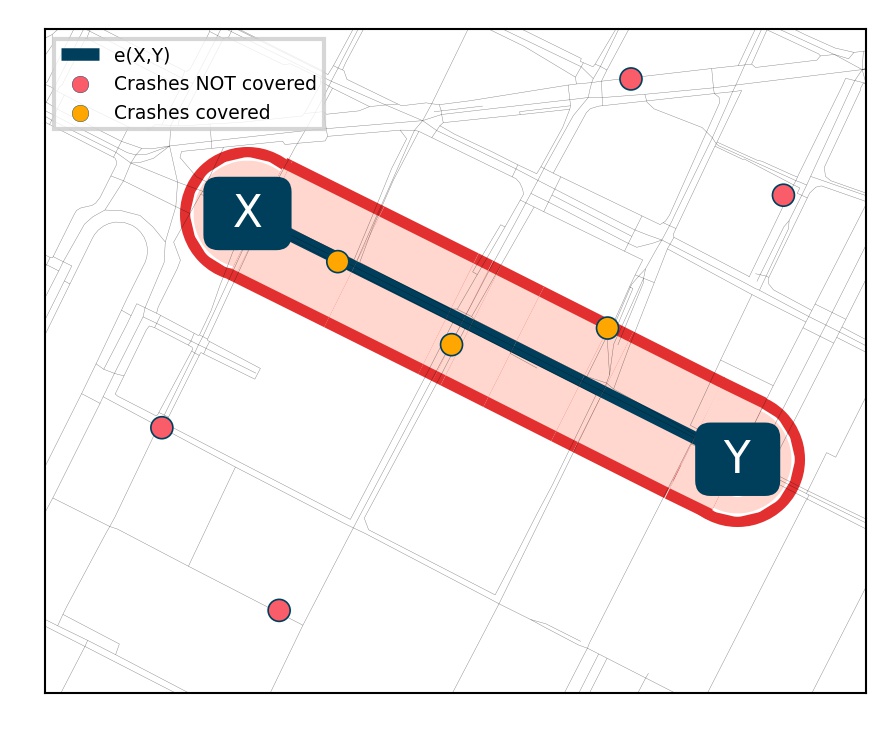}
\caption{Which crashes are covered by the link $e(X,Y)$? The red area is the area covered by link $e(X,Y)$. Any crashes within the red area will be labeled as ``covered''. We set the width of the area setting a buffer to the link. In our analysis we set this buffer $= 50\ \mathrm{m}$.} 
\label{SI_count-crashes}
\end{figure*}

\begin{figure*}[h]
\centering
\includegraphics{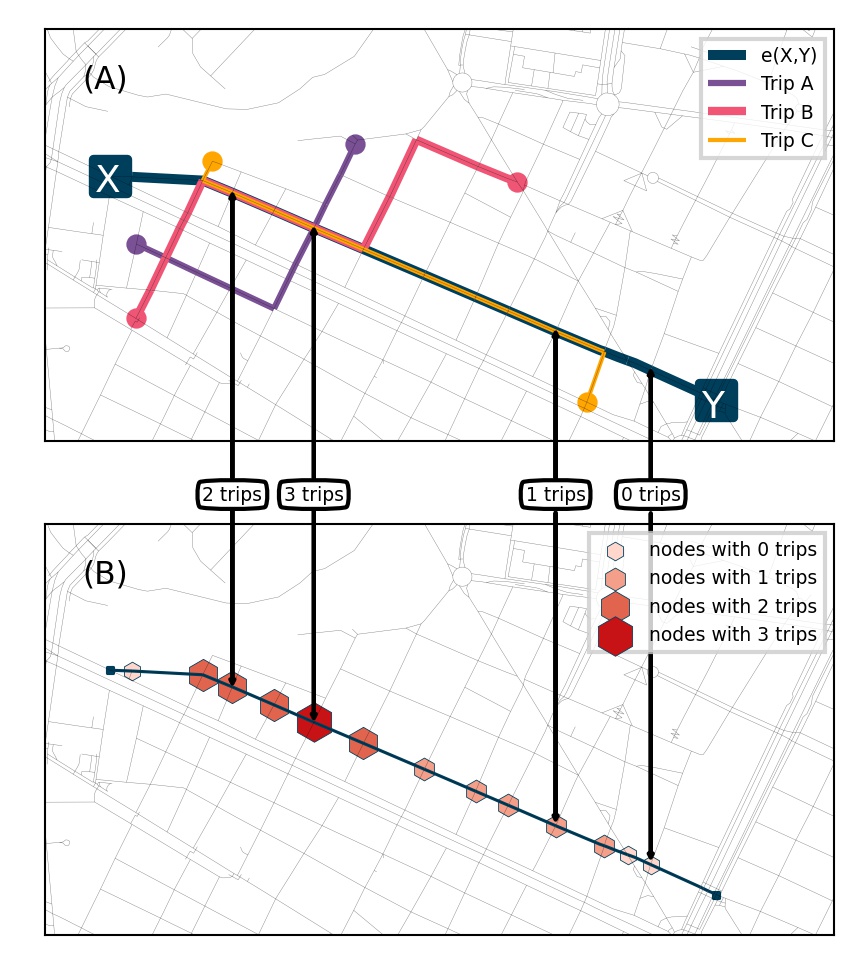}
\caption{How do we count trips on link $e(X,Y)$? In this example, we count 16 transitions on link $e(X,Y)$: $(1 \times \mathrm{Trip\ A}) + (5 \times \mathrm{Trip\ B}) + (10 \times \mathrm{Trip\ C})$. In panel \textbf{(A)} the trips are highlighted. In panel \textbf{(B)} the intersections of $e(X,Y)$ are highlighted.} 
\label{SI_count-trips}
\end{figure*}

In \textit{step 3} of the Model description section, for each link $e(X,Y)$, the weighted distance $d_{W}$ is defined. It depends not only on the distance, but also on $n_{\mathrm{crash}}$ and $n_{\mathrm{trip}}$. The variable $n_{\mathrm{crash}}$ corresponds to the number of crashes covered per $\mathrm{km}$. Where the crashes covered are all the crashes that occurred in the year 2019 less than $50\ \mathrm{m}$ away from the link $e(X,Y)$, as shown in Fig.~\ref{SI_count-crashes}. The variable $n_{\mathrm{trip}}$ counts the number of trips per $\mathrm{km}$ passing through the link $e(X,Y)$. Trips are counted as shown in Fig.~\ref{SI_count-trips}: each trip is counted a number of times equal to the number of intersections (nodes of $e(X,Y)$) crossed, in this way a different weight is given to a trip that simply crosses the link $e(X,Y)$ (see Trip A in Fig.~\ref{SI_count-trips}) compared to a trip that runs along a long part of the link $e(X,Y)$ (see Trip C in Fig.~\ref{SI_count-trips}). To compute $n_{\mathrm{trip}}$ we used $30,000$ trips sampled from our trips data set. To compute $n_{\mathrm{crash}}$ we used all the $314$ crashes of our crashes data set.

\section{Trip coverage detour = 25 \%}
In the main paper we have shown the results for detour = 0 \%, i.e. assuming that for each trip the users take the shortest path from the origin to the destination. In this way we underestimated the percentage of km traveled on the cycle path because  users are likely to agree to take a longer route if this route is safer and/or less stressful. For this reason we wanted to calculate the trip coverage also for a level of detour $> 0$ $\%$. However, how do we choose an appropriate level of detour?
In \textit{Low-Stress Bicycling and Network Connectivity} \cite{mekuria_MTI}, \textit{Mekuria et al.} define the detour level = 25 \%. They also set a condition for short trips by setting the detour level $= 0.33\ \mathrm{miles}$. In another study of nonrecreational cyclists in Vancouver \cite{vancouver_winters2010}, \textit{Winters et al.} found similar results for both bicycle and car trips: the 90 \% of trips are within 25 \% of shortest path on the road network. Finally,  \textit{Broach et al.} in their work \cite{Broach2011BicycleRC}, distinguish the results of the city of Portland between commuter cyclists and non-commuter cyclists. The former add on average 16 \% to the duration of their trip to use a cycle path, the latter add on average 26 \%.
The aforementioned studies make assumptions that we are not necessarily able to respect (for example, we do not know which of the trips we have in our dataset are made by commuter or non-commuter users). We calculate the trip coverage for 25 \% of detour.
In this way, depending on the preferences of the users (on which we do not make assumptions), the trip coverage will be between a minimum value (which correspond to the choice of the shortest route, 0 \%  of detour, see the main paper results) and a maximum value (which correspond to the 25 \% of detour accepted). To implement the detour in the model, the link length on the car network is multiplied by 1.25, then we calculate the shortest path from origin to destination. By extending the links length of the car network, the choice of links of the bike network in the calculation of the shortest path will be more likely.

\begin{figure*}[h]
\centering
\includegraphics{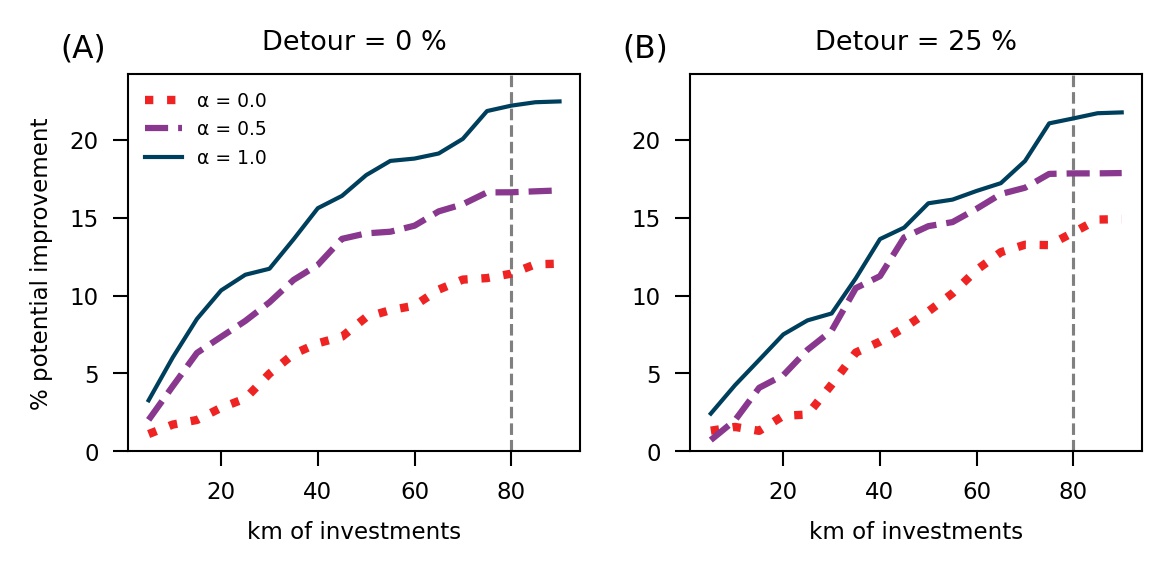}
\caption{The potential improvement of trip coverage by adding km of cycling infrastructure. \textbf{(A)} Trip coverage with 0 \% of detour for $\alpha = 0.0,0.5,1.0$. \textbf{(B)} Trip coverage with 25 \% of detour for $\alpha = 0.0,\ 0.5,\ 1.0$.} 
\label{SI_detour_results}
\end{figure*}

In Fig.~\ref{SI_detour_results} we observe the comparison of potential improvement for detour = 0 \% (panel \textbf{(A)}) and detour = 25 \% (panel \textbf{(B)}). We can observe that if detour = 25 \% there are less differences between $\alpha = 1.0$ and $\alpha = 0.5$ with respect to the results of detour = 0 \%. This is understandable because in this case the users are more flexible in choosing their route: they will be more likely to search for a cycle path to reach the destination, even if it will not exactly correspond to the shortest path. We can also observe that the potential improvements are of the same magnitude. For instance, considering $\alpha = 1.0$ at $D = 80\ \mathrm{km}$, we obtained a potential improvement of $22\ \%$ for $\mathrm{detour} = 0\ \%$ and a potential improvement of $21\ \%$ for $\mathrm{detour} = 25\ \%$.
For this we can say that the results are qualitatively the same both considering the 0 \% and 25 \% of detour. The situation is different if we consider the percentage of $\mathrm{km}$ traveled on cycle paths. For instance, considering $\alpha = 1.0$ at $D = 80\ \mathrm{km}$, the percentage of $\mathrm{km}$ traveled on the cycle paths is $29\ \%$ for $\mathrm{detour} = 0\ \%$ and $60\ \%$ for $\mathrm{detour} = 25\ \%$.
The Results tables section shows the metrics values expressed as a percentage for $\alpha = 0.0,\ 0.5,\ 1.0$ for both levels of detour mentioned.

\section{Network growth up to $D = 295\ \mathrm{km}$}
\begin{figure*}[h]
\centering
\includegraphics{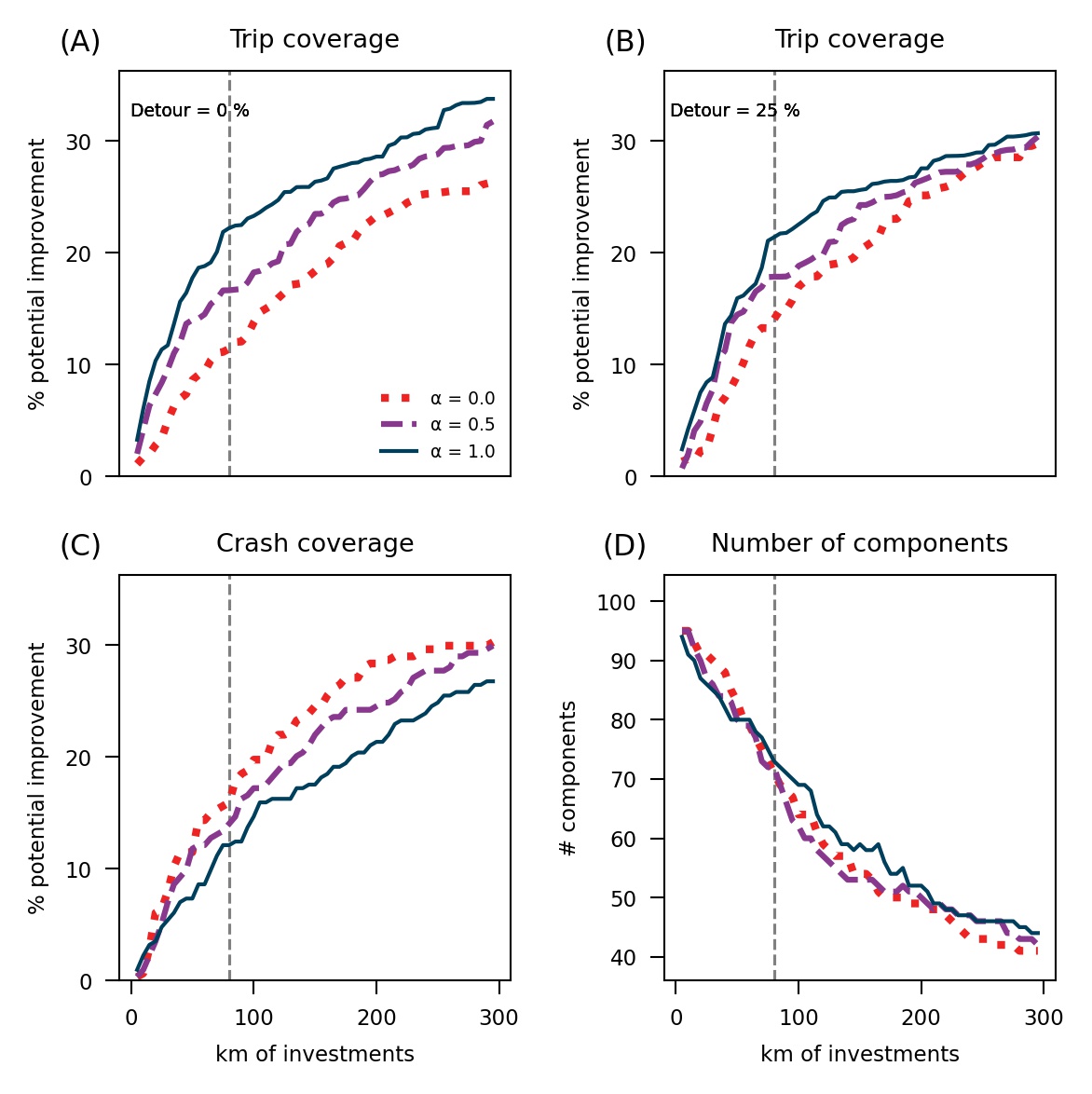}
\caption{The potential improvement of trip coverage, crash coverage and the number of components up to $295\ \mathrm{km}$. \textbf{(A)} Trip coverage with 0 \% of detour for $\alpha = 0.0,0.5,1.0$. \textbf{(B)} Trip coverage with 25 \% of detour for $\alpha = 0.0,\ 0.5,\ 1.0$. \textbf{(C)} Crash coverage for $\alpha = 0.0,\ 0.5,\ 1.0$. \textbf{(D)} Number of components for $\alpha = 0.0,\ 0.5,\ 1.0$.} 
\label{SI_300km_results}
\end{figure*}

We asked how does this model work for massive investments (i.e. very high values of $D$). For this reason, we computed the network growth up to 295 km of investments: $D = [5,295]\ \mathrm{km},\ \mathrm{step} = 5\ \mathrm{km};\ \alpha = 0.0,\ 0.5,\ 1.0$. Then we calculated the trip coverage ($\mathrm{detour} = 0\ \%,\ 25\ \%$), the crash coverage and the number of components every $5\ \mathrm{km}$ added. In Fig.~\ref{SI_300km_results} we observe the trend of these metrics. They all have in common that the potential improvement curve becomes less steep as $D$ increases, but has not yet reached the plateau. This means that, despite investments for almost 300 new $\mathrm{km}$ of infrastructure, there is still room for improvement. In Fig.~\ref{SI_300km_results}B we observe that, as already mentioned in the section Trip coverage detour = 25 \%, the potential improvements obtained for trip coverage ($\mathrm{detour}= 25 \%$) for the different $\alpha$, are more similar to each other and tend to be more and more similar as $D$ increases. Instead, we observe for trip coverage ($\mathrm{detour}= 0 \%$) and crash coverage (Fig.~\ref{SI_300km_results}A and Fig.~\ref{SI_300km_results}C respectively) that the differences between the alpha values remain even for high $D$. This means that in our model the initial choice of a more ``demand oriented'' or more ``safety oriented'' investment plan has an influence not only in the first instance, but also for large investments. In Fig.~\ref{SI_300km_results}D we see the trend of the number of components for $\alpha = 0.0,\ 0.5,\ 1.0$. As we described in the Result section, our model integrates with the existing infrastructure by adding links that connect the disconnected components. We measured the number of components of our network finding 102 components for the existing infrastructure. At  $80\ \mathrm{km}$ of investments we found 71 components for $\alpha = 0.0$ and 73 components for $\alpha = 1.0$. We also discovered that the trend of the number of components does not depend on $\alpha$: as this parameter varies, at a given $D$, the number of components remains approximately the same. We verified this by calculating the linear correlation for each pair of sequences of number of components, where each sequence corresponds to the number of components from $D = 5\ \mathrm{km}$ to $D = 90\ \mathrm{km}$ for a specific value of $\alpha$.

\section{Comparison between greedy triangulation and Delaunay triangulation}
\begin{figure*}[h]
\centering
\includegraphics{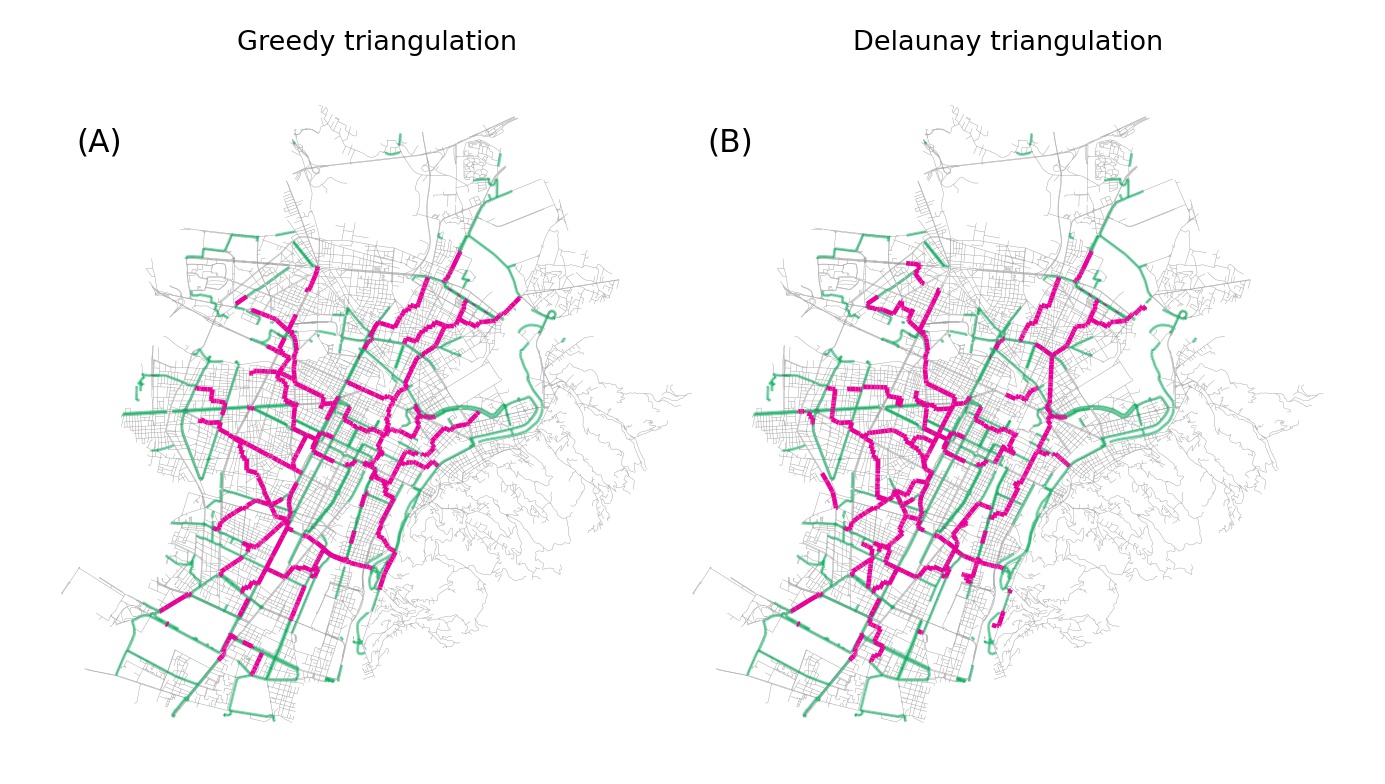}
\caption{The proposed solutions at $80\,\mathrm{km}$ of investments for $\alpha = 0.5$. using the model described in section Model description. \textbf{(A)} The proposed solution at $80\,\mathrm{km}$ of investments for $\alpha = 0.5$ using the greedy triangulation. \textbf{(B)} The proposed solution at $80\,\mathrm{km}$ of investments for $\alpha = 0.5$ using the Delaunay triangulation. The two network extensions show a few expected microscopic differences but lead to qualitatively similar improvements, see Fig.~\ref{metrics_delaunay}.} 
\label{maps_80km}
\end{figure*}

\begin{figure*}[h]
\centering
\includegraphics{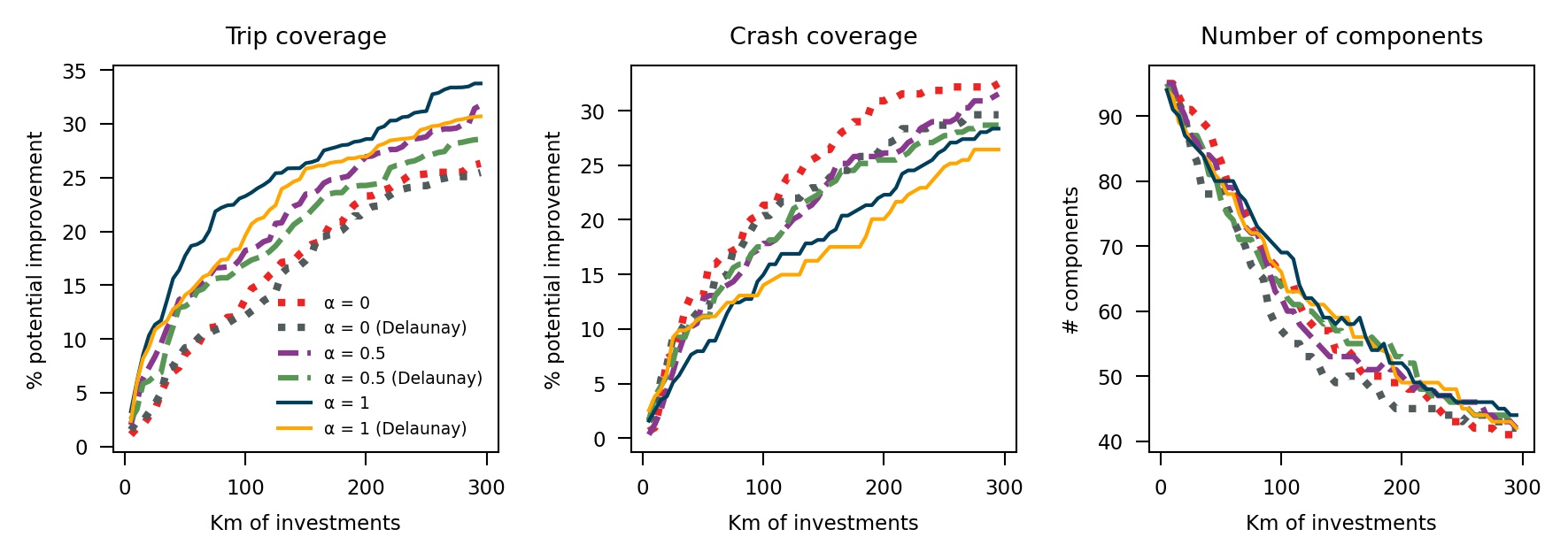}
\caption{Evaluation of the metrics Trip Coverage, Crash Coverage and Number of Components to compare the solutions obtained using the greedy triangulation and the Delaunay triangulation up to $295\ \mathrm{km}$. \textbf{(A)} Trip coverage (detour = 0 \%). \textbf{(B)} Crash coverage. \textbf{(C)} Number of components.}
\label{metrics_delaunay}
\end{figure*}

We asked how the proposed solution of our framework would change using another triangulation process, in particular when replacing the greedy triangulation with a Delaunay triangulation in our network growth model. We compared the solutions provided by our framework with these two triangulation processes computing the network growth for $D = [5,295]\ \mathrm{km},\ \mathrm{step} = 5\ \mathrm{km};\ \alpha = 0.0,\ 0.5,\ 1.0$ and $\delta = 300  \mathrm{m}$. First, for each $\alpha$, we calculated the percentage of $\mathrm{km}$ of overlap between the two solutions. The percentage of $\mathrm{km}$ of overlap does not vary much neither by varying $\alpha$, nor by varying $D$. It is on average $> 50 \%$. We reported the mean overlap values and the overlaps for $D = 80 \mathrm{km}$ in Table~\ref{T3}. In Fig.~\ref{maps_80km} we observe the proposed solutions at $D = 80 \mathrm{km}$ for $\alpha = 0.5$.
From this preliminary analysis it appears that using these two triangulation processes, the proposed solutions are similar ($> 50 \%$ overlap on average), but they are not the same. Therefore, we compared the solutions on the metrics Trip Coverage ($detour = 0 \%$), Crash Coverage, Number of Components. The results obtained are very similar on all metrics, but on average the greedy triangulation process achieves slightly better results. Only in the number of components for $\alpha = 0$ this trend is opposite and the solution using Delaunay triangulation obtained better results. The trends of metrics are shown in Fig.~\ref{metrics_delaunay}.

Note that potential boundary effects play no issue as we study growth only until 300 km of network length, and in this range it is much too early for the boundary to come into play -- the same way as the boundary links only come into play in the very last steps of the growth process in the motivating paper \cite{szell2021gub} on which we are basing our algorithm. If the city were to develop the full triangulation (beyond several hundred kilometers), boundary effects would need to be accounted for.

\begin{table}[h]
\small\sf\centering
\caption{Percentage of $\mathrm{km}$ of overlap between the solutions obtained using the greedy triangulation and the Delaunay triangulation.\label{T3}}
\begin{tabular}{|c|c|c|c|}
\hline
 $\alpha$ & 0.0 & 0.5 & 1.0 \\
\hline
\texttt{Average value} & 64 \% & 61 \% & 62 \% \\
\texttt{At $D = 80 \mathrm{km}$} & 52 \% & 57 \% & 60 \% \\
\hline
\end{tabular}\\[10pt]
\end{table}

\section{Correlation of bicycle and e-scooter trips in Milan and Rome}
One reason why we only used e-scooter trips data is that, to the best of our knowledge, bike trips data are not collected or being made available publicly in Turin. In this section we compare the data on bicycle trips with the data on e-scooter trips in two Italian cities where such data are available: Milan and Rome. We collected trip data in the same way we did for Turin (see subsection \emph{E-scooter real-time positions} in the main text for details).

For both cities we defined an e-scooter trips data set (using \emph{Bird} vehicle positions) and a bicycle trips data set (using \emph{Lime}\footnote{Lime is an e-scooter and e-bike sharing company. Access instructions to the Lime API are available at: \url{https://github.com/ubahnverleih/WoBike/blob/master/Lime.md}} vehicle positions, filtering the data to remove all vehicles that are not labeled as e-bikes). All data were collected from July 1, 2021 to November 8, 2021. To compare the e-scooter trips data with the bike trips data, for both cities, we calculated the Pearson correlation coefficient between the number of e-scooter trips per link and the number of bike trips per link (to see how we count the number of trips per link, see Section~\ref{weighted_dist}). The size of the data sets depends on the city. In Milan we randomly selected $30,\!000$ trips for each data set. In Rome we were able to define data sets with $50,\!000$ trips each, because the number of samples of the OD matrices is larger.\\
We found a high correlation between the number of e-scooter trips per link and the number of bike trips per link for both Milan and Rome: Pearson coefficient $= 0.89$ and $0.68$ respectively.

\section{Correlation of bicycle crash and e-scooter trips in Turin}
\begin{figure*}[h]
\centering
\includegraphics{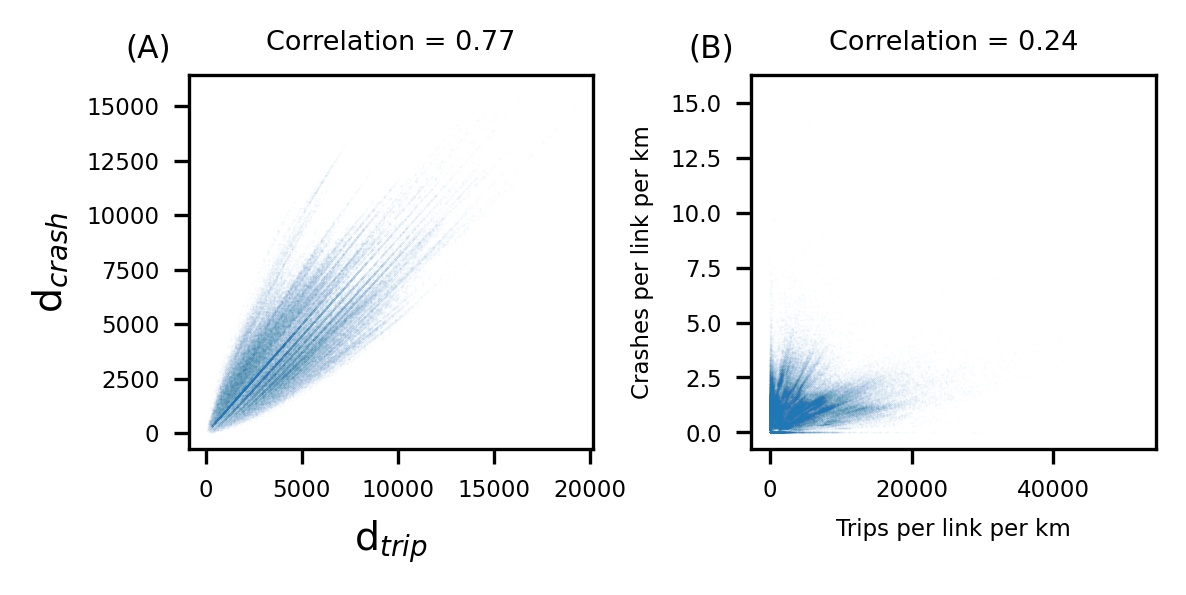}
\caption{Correlation of bicycle crash and e-scooter trips. \textbf{(A)} Correlation between $d_{\mathrm{trip}}$ and $d_{\mathrm{crash}}$. \textbf{(B)} Correlation between the \emph{number of crashes per link per km} and \emph{the number of trips per link per km}.}
\label{correlation_dtrips-vs-dcrashes}
\end{figure*}

In this section we evaluate the correlation of the two micromobility data sets we incorporated into the weighted distance $d_{W}$ that we defined in \textit{step 3} of the Model description section. For this purpose, we calculated the Pearson correlation coefficient between the quantities defined in Eq.~1: $d_{\mathrm{trip}}$ and $d_{\mathrm{crash}}$. Since $d_{\mathrm{trip}}$ and $d_{\mathrm{crash}}$ are both proportional to the length of the link, we expected a strong correlation between them. In fact, we found a Pearson coefficient $= 0.77$ (see Fig.~\ref{correlation_dtrips-vs-dcrashes}A). We also calculated the correlation between the \emph{number of crashes per link per km} and \emph{the number of trips per link per km} finding a Pearson coefficient  $= 0.24$ (see Fig.~\ref{correlation_dtrips-vs-dcrashes}B).

\section{Results tables}
The values of the metrics crash coverage, trip coverage (detour = 0 \%, 25 \%) and number of components are shown below for $\alpha = 0.0,0.5,1.0$. All the potential improvements values cited in the Results section of the main paper are obtained from these tables applying:
\begin{align}
    \pi(D) = \mu(D) - \mu_{0}
    \label{metric_improvement}
\end{align}
Where: $\mu(D)$ is the metric $\mu$ calculated with $D\ \mathrm{km}$ of new added links, $\mu_{0}$ is the metric $\mu$ calculated on the existing bicycle infrastructure (i.e. when $D = 0\ \mathrm{km}$) and $\pi(D)$ is the potential improvement of the metric $\mu$.

\begin{table}[h]
\small\sf\centering
\caption{Metrics values, $\alpha = 0.0$.\label{T0}}
\begin{tabular}{|c|c|c|c|c|}
\hline
D [km] & Crash cov. [\%] & Trip cov. (detour = 0\%) [\%] & Trip cov. (detour = 25\%) [\%] & \# components\\
\hline
\texttt{0} & 38.2 & 6.8 & 39.0 & 102\\
\texttt{5} & 38.5 & 7.5 & 40.3 & 95\\
\texttt{10} & 38.9 & 8.1 & 40.5 & 95\\
\texttt{15} & 41.1 & 8.4 & 40.3 & 93\\
\texttt{20} & 44.3 & 9.2 & 41.3 & 91\\
\texttt{25} & 44.6 & 9.8 & 41.3 & 91\\
\texttt{30} & 46.2 & 11.4 & 43.3 & 90\\
\texttt{35} & 48.4 & 12.7 & 45.3 & 89\\
\texttt{40} & 49.7 & 13.3 & 46.0 & 88\\
\texttt{45} & 49.7 & 13.8 & 46.9 & 85\\
\texttt{50} & 49.7 & 15.0 & 48.0 & 83\\
\texttt{55} & 52.5 & 15.5 & 49.1 & 80\\
\texttt{60} & 52.5 & 15.8 & 50.6 & 79\\
\texttt{65} & 53.2 & 16.8 & 51.8 & 77\\
\texttt{70} & 53.5 & 17.4 & 52.2 & 75\\
\texttt{75} & 53.8 & 17.5 & 52.2 & 73\\
\texttt{80} & 54.1 & 17.8 & 53.0 & 71\\
\texttt{85} & 56.1 & 18.4 & 53.9 & 69\\
\texttt{90} & 56.7 & 18.5 & 53.9 & 68\\
\hline
\end{tabular}\\[10pt]
\end{table}

\begin{table}[h]
\small\sf\centering
\caption{Metrics values, $\alpha = 0.5$.\label{T1}}
\begin{tabular}{|c|c|c|c|c|}
\hline
D [km] & Crash cov. [\%] & Trip cov. (detour = 0\%) [\%] & Trip cov. (detour = 25\%) [\%] & \# components\\
\hline
\texttt{0} & 38.2& 6.8 & 39.0 & 102\\
\texttt{5} & 38.5 & 8.4 & 39.7 & 95\\
\texttt{10} & 39.2 & 10.6 & 41.0 & 95\\
\texttt{15} & 40.4 & 12.7 & 43.1 & 92\\
\texttt{20} & 41.7 & 13.8 & 43.9 & 90\\
\texttt{25} & 43.3 & 14.8 & 45.5 & 87\\
\texttt{30} & 45.2 & 16.0 & 46.7 & 86\\
\texttt{35} & 46.8 & 17.4 & 49.5 & 84\\
\texttt{40} & 47.5 & 18.4 & 50.2 & 84\\
\texttt{45} & 48.1 & 20.1 & 52.7 & 83\\
\texttt{50} & 50.0 & 20.4 & 53.4 & 80\\
\texttt{55} & 50.3 & 20.5 & 53.7 & 79\\
\texttt{60} & 50.3 & 20.9 & 54.6 & 79\\
\texttt{65} & 51.0 & 21.8 & 55.5 & 77\\
\texttt{70} & 51.3 & 22.3 & 55.9 & 73\\
\texttt{75} & 51.6 & 23.0 & 56.8 & 72\\
\texttt{80} & 52.2 & 23.0 & 56.8 & 72\\
\texttt{85} & 52.9 & 23.1 & 56.8 & 69\\
\texttt{90} & 54.5 & 23.2 & 56.9 & 66\\
\hline
\end{tabular}\\[10pt]
\end{table}

\begin{table}[h]
\small\sf\centering
\caption{Metrics values, $\alpha = 1.0$.\label{T2}}
\begin{tabular}{|c|c|c|c|c|}
\hline
D [km] & Crash cov. [\%] & Trip cov. (detour = 0\%) [\%] & Trip cov. (detour = 25\%) [\%] & \# components\\
\hline
\texttt{0} & 38.2& 6.8 & 39.0 & 102\\
\texttt{5} & 39.2 & 9.7 & 41.4 & 94\\
\texttt{10} & 40.4 & 12.4 & 43.2 & 91\\
\texttt{15} & 41.4 & 14.9 & 44.8 & 90\\
\texttt{20} & 41.7 & 16.7 & 46.5 & 87\\
\texttt{25} & 43.0 & 17.7 & 47.4 & 86\\
\texttt{30} & 43.6 & 18.1 & 47.8 & 85\\
\texttt{35} & 44.3 & 20.0 & 50.1 & 84\\
\texttt{40} & 45.2 & 22.0 & 52.6 & 82\\
\texttt{45} & 45.5 & 22.8 & 53.4 & 80\\
\texttt{50} & 45.5 & 24.1 & 54.9 & 80\\
\texttt{55} & 46.8 & 25.1 & 55.1 & 80\\
\texttt{60} & 46.8 & 25.2 & 55.7 & 80\\
\texttt{65} & 48.1 & 25.5 & 56.2 & 78\\
\texttt{70} & 49.4 & 26.5 & 57.6 & 77\\
\texttt{75} & 50.3 & 28.3 & 60.0 & 75\\
\texttt{80} & 50.3 & 28.6 & 60.4 & 73\\
\texttt{85} & 50.6 & 28.8 & 60.7 & 72\\
\texttt{90} & 50.6 & 28.9 & 60.8 & 71\\
\hline
\end{tabular}\\[10pt]
\end{table}

\clearpage


\bibliography{main}
\bibliographystyle{SageV}